# Perytons and their Possible Sources

Mohammad Danish Khan
Bombardier Aerospace
123 Garratt Blvd., Toronto, ON, M3K1Y5, Canada

**ABSTRACT**

Perytons are terrestrial signals that exhibit dispersion measure (DM) similar to pulsars. In trying to identify terrestrial sources of such perytons, investigation into signals from airborne equipment (aircraft), RFI emissions from electronics and lightning phenomenon reveals that the possible sources of perytons could be lightning phenomenon. Narrow Bipolar Pulses (NBPs) and Terrestrial Gamma Flashes (TGFs) are good investigational candidates.

## 1. Introduction

In trying to determine possible sources of Perytons[1] (mysterious signals that although exhibit the 'traversed free electron column density' also known as 'dispersion measure' (DM) similar to Fast Radio Bursts[2] (FRBs) or Lorimer Burst[3] (LB) or pulsars, have symmetric pulse width in 10s of milli-seconds (approx. 30 – 50 ms) (S. Burke-Spolaor et al.) and are determined to be of terrestrial origin) information detailing characteristics of Perytons and their similarities and differences from FRBs or LB were researched into. Below are some of the differentiating characteristics between Perytons and FRBs/LB:

| Characteristics | Perytons | FRBs/LB |
|---|---|---|
| 1 | Multi-Bean Detection (detected in all 13 beams of the 20cm multi-beam receiver of Parkes Telescope in NSW, Australia). So far Perytons have been detected only at Parkes (Australia) and Bleien Observatory (Switzerland) | LB was detected in 3 of the 13 beams of the multi-beam receiver of Parkes telescope, NSW, Australia. |
| 2 | Symmetric Pulse Profile | Asymmetric Pulse Profile |
| 3 | 30-50ms (10s of milli-seconds) duration pulses (pulse width) in the 1.1 to 1.5 GHz range and peak flux density from 50 – 300 mJy (milli-Jansky) | Pulse width is much shorter than Perytons (5 ms duration and high intensity of 30 Jy for LB in 1.4 GHz band and less than 10ms pulses for FRBs) |
| 4 | Arrival time of pulses is inversely proportional to $f^k$ where $k \approx 2$ although some perytons have shown k=2 characteristics within the experimental errors (S.Burke-Spolaor et al.) | Arrival time of pulses is inversely proportional to $f^k$ where k = 2 |
| 5 | Although some perytons exhibit deviation from dispersive delay mode (S.Burke-Spolaor et al.), DM is in the range of 200-400 $cm^{-3}$ pc (mode at $\approx$ 380 $cm^{-3}$ pc) | DM in the case of LB was 375 $cm^{-3}$ pc and for FRBs varied from 553-1104 $cm^{-3}$ pc. |

Table 1

1. Perytons are mythological creatures with physical characteristics of stag and bird casting the shadow of man. Since the Perytons in this context of the presentation, are terrestrial pulses that mimic the characteristics of extra-galactic pulses, they have being names "Perytons" by their discoverer (S.Burke-Spolaor et al. 2011)
2. FRB (Fast Radio Busts) Thornton et al. 2013
3. Lorimer Burst (LB) L.B. Lorimer et al. 2007

## 2. Possible Sources of Perytons

In order to find the possible sources of Perytons, several papers (referenced throughout this note) were reviewed and the following was inferred:

1. All the perytons detection occurred during day time (with the exception of one at Bleien Observatory (P. Saint-Hilaire et al. 2014)).
2. At Parkes, eleven signals were detected in a single 4.4 minute interval observational period.
3. 4 of the 5 signal detections at Parkes occurred in June/July period through years 1998-2003 (mid-winter time period in Australia). These detections were non-random on an annually and daily prediction basis (S.Burke-Spolaor et al.)
4. At Parkes, 15 independent perytons detection were observed and at Bleien 5 independent perytons detection were observed (one of 5 perytons detected at Bleien was detected at night-time)
5. At Parkes, signals lacked regular periodicity and were continuous emissions in the 1.4 GHz band (S.Burke-Spolaor et al.). Moreover, Kocz et al. (2012) on re-analysis noted train of perytons separated by 22s. Therefore it is reasonable to infer from 2) and 5) that some of the perytons occurred closely spaced in time.
6. It is then clear that perytons are either made-made or local natural phenomenon and although mimic DM of pulsars are certainly terrestrial in origin.
7. So far, most of the detections were observed at Parkes (15 were observed at Parkes and 5 at Bleien). Therefore majority (not all) of the discussions on possible sources are based on these observations.

With all this information, we now look at the possible sources of Perytons. The below are found to be **possible sources of perytons** and are discussed in detail that follows:

a) Lightning Events in the ionosphere/Solar Bursts (Natural high-altitude phenomenon).
b) Signals from aircraft to ground based stations (navigation and communication) and from GPS satellites to ground based and airborne receivers (including aircraft).
c) Emissive signals from electronics.
d) Multiple Bursts which is a lightning phenomenon predominately occurring as inter-cloud and intra-cloud electric discharge. Narrow Bipolar Pulses (NBPs) and Terrestrial Gamma Ray Flashes (TGFs).

### A. Lightning Events in the ionosphere/Solar Bursts (Natural high-altitude phenomenon)

Based on perytons detected at Parkes, the perytons detections was predictable on annual cycles (June/July period) between years 1998-2003 and these detections were non-random (occurrence time-wise) (S.Burke-Spolaor et al.).

June/July period is a typical peak period of thunderstorm activity in that (Australia) part of the world. For example, South China Sea and nearby area is popular (in Aviation Industry) for weather activity during this time and transient luminous events are common in the lower altitude (troposphere). However, transient luminous events between the thunderclouds and ionosphere have been discovered as well (see for instance H.T.Su et al. 'Gigantic Jets between a thundercloud and the ionosphere' Nature. Volume No. 423). In this activity, thunderstorms actually serve as generators driving electric current upwards from higher altitude cloud-tops to ionosphere (HT. Su et al.) Although, this lightning phenomenon is of low frequency nature, the progressive incitement of plasma oscillations in regions of varying plasma density can potentially lead to non-dispersive, swept emissions at sweep rate in the range of approx. 1 GHz/s. For example type III solar burst along with this lightning event (between higher altitude clouds and ionosphere) could have caused emissions in the observation band at Parkes (S.Burke-Spolaor et al.).

Other lightning phenomenon worth mentioning here is 'Trichel Pulses'. These pulses have current rise time of between 25-50ns and half values are attained in twice that time. These pulses are short and can lead to radio noise over a very wide band of frequencies (Fisher et al.) and have been reported from 2 KHz to well above 3 MHz. The phenomenon of lightning flashes (negative first stroke) has been recorded to peak at above 250 Kilo-Amps with time to half values at about 2-3 milli-seconds.

Moreover, at Bleien observatory, perytons were either observed during summer or winter times only (1 in July, 2 on Oct, 1 in Nov and one observed during night time in December) which also happens to be a period of significant weather activity in central Europe. Before (4.5 minutes) the observation of one of the perytons at Bleien (summer-time Perytons observation), an 'atmospheric discharge' was visible in the solar observations from 20-80MHz as well as winter-time Perytons observation was followed by 'a broadband short but not drifting event' (P. Saint-Hilaire et al. 2014).

Therefore, it is possible given the fact that majority of perytons detection events were recorded during a certain (predictable) time of the year that these events are caused by natural terrestrial source than man-made signals.

### B. Signals from aircraft to ground based stations (navigation and communication) and from GPS satellites to ground based and airborne receivers (including aircraft)

Below are the most common transmit and receive systems, along with their frequency range and polarization aboard modern aircraft from frequency range of 1.1 GHz to 1.5 GHz and as detailed:

1. **GPS L1** (Receive System only):
   Frequency Range: 1.565 GHz – 1.585 GHz
   Polarization: RHCP (Right Hand Circularly Polarized)
   Characteristics: Based on frequency allocation filing (with L1 Coarse Acquisition (C/A) power as 26 dBw and antenna gain as 13 dBi), the power is about 500W (27 dBw). With the 21,000 Km free space path loss (from satellite to ground station) of 182 dB, we have received power in the range of a few femto ($10^{-15}$) watts (-155 dB or -125 dBm). C/A (pseudo random noise) code is modulated onto the L1 carrier only.
   Pulse width of signals is typically of the order of 10s of nano-seconds and the pulse repeats depends on the signal ranging code (for example C/A, code repeats every 1ms and P code every week).

2. **GPS L2** (Receive System only):
   Frequency Range: 1.217 GHz – 1.238 GHz
   Polarization: RHCP (Right Hand Circularly Polarized)

Characteristics: Same characteristics as GPS L1 except that the received power for unity gain RHCP antenna is -160 dB or -130 dBm).
Pulse width of signals is typically of the order of 10s of nano-seconds and the pulse repeats depends on the signal ranging code (for example C/A, code repeats every 1ms and P code every week).

3. **GLONASS** (Russian/Receive System only):
   Frequency Range: 1.240 GHz – 1.256 GHz
   Polarization: RHCP (Right Hand Circularly Polarized)
   Characteristics: Message capacity of GLONASS is significantly lower than GPS. Although most characteristics of GLONASS are similar to that of GPS, it only uses L2 primary frequency channels and unlike GPS, GLONASS system transmits the same code on different frequencies.
   Pulse width of signals is typically of the order of a few micro-seconds.

4. **Inmarsat Satcom Civilian Downlink** (Receive System only):
   Frequency Range: 1.53 GHz – 1.56 GHz
   Polarization: All
   Characteristics: The characteristics of INMARSAT Satcom system onboard a commercial aircraft depends on the system type itself but typically the receiver has maximum and minimum input signal level of -17dBm (assuming a maximally loaded satellite signal of -78.20 dBm and a maximum interference level of -72 dBm into the LNA. Also assumed is a maximum LNA gain of 60 dB and a minimum RFU-LNA cable loss of 6.00 dB) and -71 dBm respectively.
   Pulse width of signals is 10 micro-seconds.

5. **SSR** (Secondary Surveillance Radar):
   Frequency Range: 1.03 GHz – 1.09 GHz
   Polarization: Vertical
   Characteristics: Requires radar transponder on the aircraft and requests an aircraft's attitude and identity by Air Traffic Controllers for airspace surveillance. This system is based on IFF (Identification Friend or Foe) technology.
   Pulse width of signals is typically tenths of microseconds. Pulse spacing ranges from 3 – 20 microseconds. Perytons being broad pulses with broad band emissions preclude them for being of a radar origin.

6. **TCAS** (Traffic Collision Avoidance System):
    Frequency Range: 1.03 GHz – 1.09 GHz
   Polarization: Vertical
   Characteristics: TCAS is an airborne electronic system that employs radio signals for surveillance (similar to SSR) of nearby aircraft and can display and generate alarms (audible) in the cockpit to alert pilots. The interrogations are transmitted once per second. Nominal power level for transponder is 250W radiated and nominal receiver sensitivity is -74dBm.
   Pulse width of signals is typically 0.5 microseconds.

To add to the above, Flight Test Vehicles (FTVs) typically utilize on board telemetry systems which commonly use frequency in the 1.4 GHz band. The typical transmitted RF power is 10 W (40 dBm) with a bandwidth of 9 MHz.

As can be realized from information above for different aircraft systems utilizing frequency range from 1.1 GHz to 1.5 GHz, that 30-50ms pulse width of perytons is not a characteristic of communication and navigation systems on board modern commercial aircraft (this excludes military and specialized mission aircraft).

Moreover at Parkes, detection of 11 Perytons event in one 4.4 minute duration can exclude aircraft signal origin, since the aircraft would have been well out of range within a much shorter time than 4.4 minutes for the Parkes telescope to detect signals from it (assuming aircraft cruising ground speed of typically 700 – 900 Km/h, it would have travelled about 30 Km within half of this 4.4 minutes). It is possible that some of the satellite signals can exhibit some dispersion (encounter dispersive medium) which remains to be analyzed, however perytons having broad pulse characteristics (10s of milliseconds) makes it difficult for a convincing argument for perytons to have originated as a commercial aircraft or satellite signal of systems described above with pulse width of a few micro-seconds.

### C. Emissive Signals from Electronics

1.4 GHz radio frequency band is regulated by FCC (Federal Communications Commission) from unwanted emissions produced by electronic devices (under Title 47, Title 15 of Code of federal Regulations) in the US and the same regulations are observed worldwide by appropriate regional regulatory agencies.

Both Class A and B devices (commercial and residential) are regulated and require devices to be tested against having certain maximum field strength typically within a few meters. Therefore emissions from an electronic device with sufficient intensity (resulting in a perytons observation) suggest that this emissive device must have been located on site at Parkes Observatory. However, S. Burke-Spolaor et al. (2011) argue against an emissive electronic device at Parkes on empirical grounds and rule out emissions from a hardware failure of electronic device onsite.

Moreover, since the perytons detection at Parkes and Bleien followed a day-time detection trend as well as in case of Parkes followed semi-annual cycle, the emissions from a malfunctioning electronic device can be ruled out.

### D. Multiple Bursts (lightning phenomenon) predominately occurring as intra-cloud electric discharge. NBPs and TGFs.

Intra cloud lightning pulses (inter and intra cloud lightning is often referred to as multiple bursts) are a lightning phenomenon that involves large number of relatively low amplitude pulses (current) which are of short duration. These normally occur at the later stage of intra-cloud lightning flashes (M.A. Uman et al.). Research in this area (lightning) has been somewhat limited to Power Industry and to a lesser degree to Aerospace Industry.

Within the multiple bursts environment, Narrow Bipolar Pulses (NBP) would be of much interest in the context of this presentation (perytons) since they are characterized by the following:

1. Narrow Bipolar Pulses are higher altitude discharges with altitude location for negative discharges (transports negative charges downwards) between 15 Km to 20 Km and altitude location for positive discharges between 7 Km – 14 Km.
2. Almost 'No optical' emissions.
3. Accompanied by strong RF emissions (observed up to 30 MHz)
4. Short pulse duration (typically 10s of micro-seconds).
5. Occur as isolated events without followed or preceded by other lightning activity for at least tens of milli-seconds.
6. More commonly observed in the tropical regions than temperate regions, however due to the limited research in this field, this has not been verified.
7. Thought to be associated with the initiation of intra-cloud discharges (Jacobson et al.)

Other forms of lightning such as inter-cloud flashes can also be a source of RF emissions, NBPs show much stronger RF emissions when compared with other types of lightning flashes and their practically no optical emissions makes them a candidate of interest here. Although the paper on the perytons observation at Bleien concludes that no lightning was observed on site at the time of the events, it is possible due to 'non-optical' emissions nature of NBPs that these flashes did indeed occur but were not visible.

The observations/study of NBPs was primarily done by satellite based or broadband electric field measuring systems (N.A. Ahmad 2011). Below is an example of NBP:

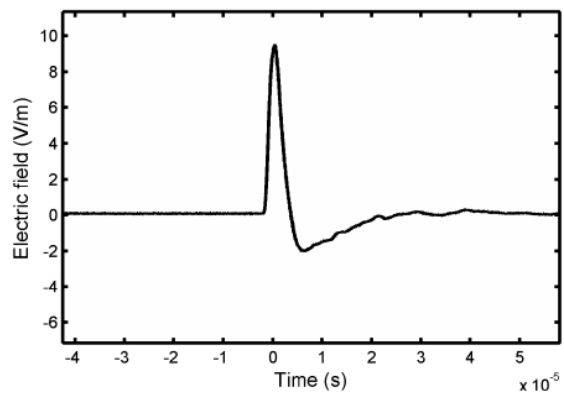

Figure 1 (Narrow Positive Bipolar Pulse) (Courtesy: N.A. Ahmad, 2011)

Although the typical duration (pulse width) of NBPs has been in the vicinity of tens of micro-seconds, longer duration pulses have also been reported (Villaneuva et al.) and High Frequency emissions of 30 MHz has been observed associated with NBPs. It is possible that NBPs associated with other natural phenomenon was recorded as a Perryton event at Bleien or Parkes observatory.

Lightning phenomenon such as initial breakdown in cloud flashes has been reported to be associated with pulses of up to 80 micro-seconds with almost milli-second duration between pulses (Rakov et al.), this phenomenon in general has not been well studied and requires more research work. Multiple burst phenomenon has primarily been studied using instrumented aircraft specifically 'prepared' to observe and record these events. Therefore the observations have been limited and more research is needed to better understand the nature of lightning environment including NBPs.

Another possible natural source could be Terrestrial Gamma Ray Flashes (TGFs). TGFs typically are/have (see: B.E. Carlson et. al. (Terrestrial Gamma Ray flash production by lightning)):

1. Terrestrial in Origin.
2. Approximately 1 milli-second pulse duration.
3. Associated with lightning (lightning phenomenon within 2 ms of TGFs).
4. Photons with energy of up to 20 MeV.
5. Frequency of 'at least' 50/day (although this number could be much bigger).
6. High average energy per photon than any gamma ray phenomenon from space (extra-terrestrial).

The average rate of TGFs based on RHESSI data has been about 50/day but one model predicts a global rate to be about 5000/day TGF events (H.J. Christian et. al., 2005). However, if TGFs are associated with lightning then with a reported average rate of 3-4 million/day lightning events (Uman et. al), this number of 5000/day TGF events is about 0.1% of the average daily rate of lightning events and a much larger than 5000/day TGF events rate cannot be ruled out as they may be outside of RHESSI's detection threshold (D.M. Smith et. al.). The incredible all-sky rate of FRBs as $10^4$ per day (S.R. Kulkarni et. al., 2014) compared to a possible similarly high rate of TGF events/day can possibly provide a correlation between the two.

### 3. Discussion

Perytons observations so far have been limited to Parkes and Bleien Observatories indicating that this is a global phenomenon. If it is indeed the case, then it is a matter of time before more observations are reported and more information on the nature of perytons is determined. From the observations so far of Perytons (majority of them at Parkes) it is reasonable to assume that Perytons are possibly a result of natural (terrestrial) phenomenon (lightning intra cloud flashes, NBPs along with other events). At the same time, it is also possible given the complex nature of communications on earth (through satellites and airborne equipment) that the source of perytons may be man-made, although it seems more unlikely. More observations need to be reported and researched into, to better understand and narrow down on these sources of perytons in support of radio astronomy.

The main issue to be able to narrow down the source(s) of perytons is the dispersion. Considering pulse duration from aircraft navigation systems as well as from lightning, for any of these pulses to be detected as a peryton, these pulses must chirp down over 100s of MHz in 10's of milli-seconds (very unusual for aircraft or lightning signals). Therefore, further investigation into the possibility of dense plasma associated with TGFs or NBPs is required as that could provide an explanation for the high dispersion required of these lightning pulses for them to be sources of perytons.

### References

S.R. Kulkarni et al., 'Giant Sparks at Cosmological Distances'
S-Burke-Spolaor et al., 'Radio Bursts with Extragalactic spectral characteristics show terrestrial origins'
H.T Su et al., 'Gigantic jets between a thundercloud and the ionosphere'
Fisher et al., 'Lightning Protection of Aircraft'
N.A. Ahmad, 'Broadband and HR radiation from cloud flashes and Narrow Bipolar pulses'
Villanueva et al., Microsecond scale electric field pulses in lightning discharge flashes'
P. Saint-Hilaire at al., 'Short duration radio bursts with apparent extra galactic dispersion'
V.A. Rakov et al., 'Bursts of pulses in lightning electromagnetic Radiation'
SAT 2000/2100, Rockwell Collins Inc., Cedar Rapids, Iowa
B.E. Carlson et. al., 'Terrestrial gamma ray flash production by lightning current pulses'
Fishman at. Al., 'Discovery of intense gamma ray flashes of atmospheric origin'
Inan U.S. et. al., 'Production of terrestrial gamma ray flashes by an electromagnetic pulse from lightning return stroke'